\documentclass[reprint,superscriptaddress,amsmath, amssymb,aps,prd,notitlepage,longbibliography,floatfix,nofootinbib,onecolumn]{revtex4-1}

\usepackage{siunitx}
\usepackage{graphicx}
\usepackage{multirow,booktabs}

\newcommand{\Eq}[1]{Eq.~\eqref{#1}} 
\newcommand{\Fig}[1]{Fig.~\ref{#1}}
\newcommand{\Table}[1]{Table~\ref{#1}}
\newcommand{\Appd}[1]{Appendix~\ref{#1}}
\newcommand{\vecn}{\boldsymbol{n}}
\newcommand{\dif}{\mathrm{d}}
\newcommand{\gw}{\mathrm{gw}}

\begin{document}
\title{Isotropic stochastic gravitational wave background reconstruction for Taiji constellation}
\author{Yang Jiang}
\email{jiangy@ucas.ac.cn}
\affiliation{School of Fundamental Physics and Mathematical Sciences, 
  Hangzhou Institute for Advanced Study, UCAS, Hangzhou 310024, China}
\affiliation{School of Physical Sciences, 
    University of Chinese Academy of Sciences, 
    No. 19A Yuquan Road, Beijing 100049, China}
\author{Qing-Guo Huang}
\email{corresponding author: huangqg@itp.ac.cn}
\affiliation{School of Fundamental Physics and Mathematical Sciences, Hangzhou Institute for Advanced Study, UCAS, Hangzhou 310024, China}
\affiliation{School of Physical Sciences, 
    University of Chinese Academy of Sciences, 
    No. 19A Yuquan Road, Beijing 100049, China}
\affiliation{Institute of Theoretical Physics, Chinese Academy of Sciences,Beijing 100190, China}

\date{\today}

\begin{abstract}
The stochastic gravitational wave background is a broadband target from diverse astrophysical and cosmological sources. The background falls within the mHz frequency band could become a potential observable for future space-based interferometers. Taiji, a proposed space mission slated for launch in the 2030s, will enable the study of such a background. However, the unique characteristics of space missions pose distinctive challenges for separating the stochastic background from instrumental noise. To address the data analysis requirements, we develop a preliminary pipeline to search for the SGWB and evaluate its performance with Taiji simulation datasets. At present, we demonstrate that the algorithm can successfully recover the parameters of injected background with a known spectral density after setting aside the complication of galactic binaries foreground. Furthermore, by employing the trans-dimensional Markov Chain Monte Carlo method, we extend the analysis to reconstruct the background with unknown spectral morphology.
\end{abstract}

\maketitle

\section{Introduction}
Over the past decade, ground-based detectors such as LIGO \cite{LIGOScientific:2014pky}, Virgo \cite{VIRGO:2014yos} and KAGRA \cite{KAGRA:2020tym} 
have achieved remarkable success, pioneering the field of gravitational wave (GW) astronomy.
By the recent O4a observing run \cite{LIGOScientific:2025slb},
more than two hundred compact binary coalescences (CBCs) have been recorded.
These GW signals are concentrated in the frequency band of several tens of hertz, 
corresponding to the systems of astronomical binary black holes, black hole-neutron star, and binary neutron stars.
Unfortunately, environmental noise, particularly seismic noise, limits the sensitivity of ground detectors
in the low-frequency regime \cite{LIGO:2024kkz}. 
To overcome this limitation, it is essential to increase the arm length and deploy the detector in space.
After long-term preparation, Laser Interferometer Space Antenna (LISA) \cite{Baker:2019nia,LISA:2024hlh}, 
a space-borne GW detecting project led by
the European Space Agency has now come into the engineering phase, 
marking a significant step toward addressing this challenge.

Cocurrently, China is also advancing space-based GW detection, scheduled for launch in the 2030s.
Two major projects have been proposed.
TianQin \cite{Luo:2025sos,Luo:2025ewp} consists of three Earth orbiting satellites, forming a constellation centered around Earth.
In contrast, Taiji is a space-borne GW observatory sponsored by Chinese Academy of Sciences,
similar in design to LISA. 
It comprises three spacecrafts (SCs) configured in an equilateral triangle constellation with a nominal arm length of $3\times 10^9$ m.
The barycenter follows a heliocentric orbit, trailing or leading the Earth by \ang{20} approximately \cite{10.1093/nsr/nwx116,10.1093/ptep/ptaa083}. 
The Taiji-1 pilot satellite, launched in 2019,
has successfully completed its mission, validating a number of key technologies \cite{doi:10.1142/S0217751X21400042,taiji1_inertial_sensor,taiji1_micro_thruster,taiji1_drag_free}
and demonstrating the feasibility of the Taiji roadmap.
These forthcoming observatories will emerge the millihertz GW sky, 
providing unprecedented opportunities for studying the GW signals from
massive black hole binaries \cite{Li_2022},
extreme mass-ratio inspirals \cite{Jonathan_2004},
Galactic binaries \cite{10.1093/mnrasl/slab003} and inspiral stellar-mass black hole binaries \cite{Buscicchio:2024asl}.

In addition to individual GW events previously discussed, the stochastic GW background (SGWB) represents another type of signal
within great scientific  interest. It typically arises from the superposition of numerous weak and unresolved sources.
Such a guaranteed source in the millihertz band is the confusion foreground from the vast population of stellar compact binaries in
the Galaxies \cite{10.1093/mnras/sty2035}. It is estimated that our galaxy contains $\mathcal{O}(10^7)$ such systems, of which only about $\mathcal{O}(10^4)$
are expected to be resolved individually \cite{Breivik:2017jip,Ruiter:2007xx}. The unresolved remainder forms a persistent GW background \cite{Liu:2023qap}.
Detection of this background will serve as a supplement to individual binaries, providing valuable insights to the structure of our Milky Way.
Furthermore, cosmological SGWB is also predicted to be within Taiji's observational band. 
Several well-motivated physics beyond the Standard Model \cite{Croon:2019kpe,Huber:2015znp,PhysRevLett.105.041301} predict a first-order phase transition (PT) \cite{Linde_1979} in early Universe.
Taiji is particularly sensitive to the phase transition at electroweak scale \cite{PhysRevD.111.L051702,Huang_2025}.
A detection of such GW background would constitute a major scientific breakthrough.
As a candidate of dark matter, the primordial black holes (PBHs) may be originated by comoving curvature pertubations
generated when inflation reenters the horizon \cite{Ivanov:1997ia}. By inspecting the accompanying GW radiation, 
Taiji has the potential to probe PBHs in the mass range of $10^{-16}\sim 10^{-10}\,M_\odot$ \cite{PhysRevD.99.103521,Yuan:2019wwo},
offering a unique window to test PBH hypothesis. 

The SGWB manifests a stochastic signal in the output of detector, akin to instrumental noise.
Distinguishing GW from noise and extracting its physical implications are the duties of data analysis technique.
This task is particularly complex for space-based interferometers.
The standard cross-correlation method, used for SGWB searches in ground detectors \cite{10.1093/mnras/227.4.933}, relies on the assumption 
that environmental noise is uncorrelated between different detectors, 
which does not hold for a single space interferometer like Taiji.
In general, detecting the SGWB in such a mission necessitates prior knowledge of the characteristics about noise.
Instrument noise originates from diverse sources \cite{10.1093/ptep/ptaa083,Barke:2015srr,PhysRevD.111.043048,sktcxb-10-3-fangzimuo}.
For demonstration purposes, current data processing strategies typically model several dominant noise components.
Earlier studies and simulations \cite{PhysRevD.82.022002,PhysRevD.89.022001} have demonstrated the feasibility of SGWB identification.
In light of recent advancements in the LISA design, studies \cite{Caprini:2019pxz,Flauger:2020qyi,Kume:2024xvh} have employed an automated adaptive frequency-binning procedure for reconstructing SGWB with arbitrary spectral shapes.
On the other hand, the data stream of a space mission will be signal-rich, containing a large number of
concurrent GW sources. Therefore, any searching program designed for particular GW signal should be integrated into a global fitting pipeline \cite{Littenberg:2023xpl,Rosati:2024lcs}. 
To address these multifaceted challenges, the Taiji Data Challenge II (TDC II) was recently released \cite{Du:2025xdq}.
The datasets incorporate the GW sources expected within Taiji's sensitivity band, and are divided 
into blind and training sets to foster the development of robust data analysis techniques.

In this work, we will concentrate on the training sets designed for SGWB science. 
Going beyond the commonly adopted equal-arm approximation, our analysis incorporates the more
realistic, simulated orbits provided in TDC II, which result in unequal and time-varying interferometric armlengths.
Furthermore, unlike the assumptions in \cite{Baghi:2023qnq}, the single-link interferometric noise levels are different across the links.
We begin by conducting a template-based parameter estimations for the background, assuming a known spectral shape.
In general, the energy spectrum of the SGWB is not a known prior.
Therefore, we choose to employ a trans-dimensional Markov chain Monte Carlo (MCMC) \cite{10.1093/biomet/82.4.711} approach
to reconstruct the frequency profile of SGWB.
This model-agnostic approach enables the detection of SGWB without the precise knowledge about the spectrum.
We compare its performance directly with that of the template-based method.

The TDC II datasets contain both isotropic SGWB and confusion galactic binaries (CGBs).
The latter constitutes an anisotropic background due to the inhomogeneous distribution of binaries in the Milky Way,
exhibiting a year-round modulation pattern. We postpone the issue of CGB to future work. 
At this stage, we pre-process the dataset by subtracting the provided CGB component and 
focus our analysis on the residual noise and isotropic background.

\section{TDC II datasets}
\begin{figure}[ht]
  \centering
  \includegraphics[width=.6\columnwidth]{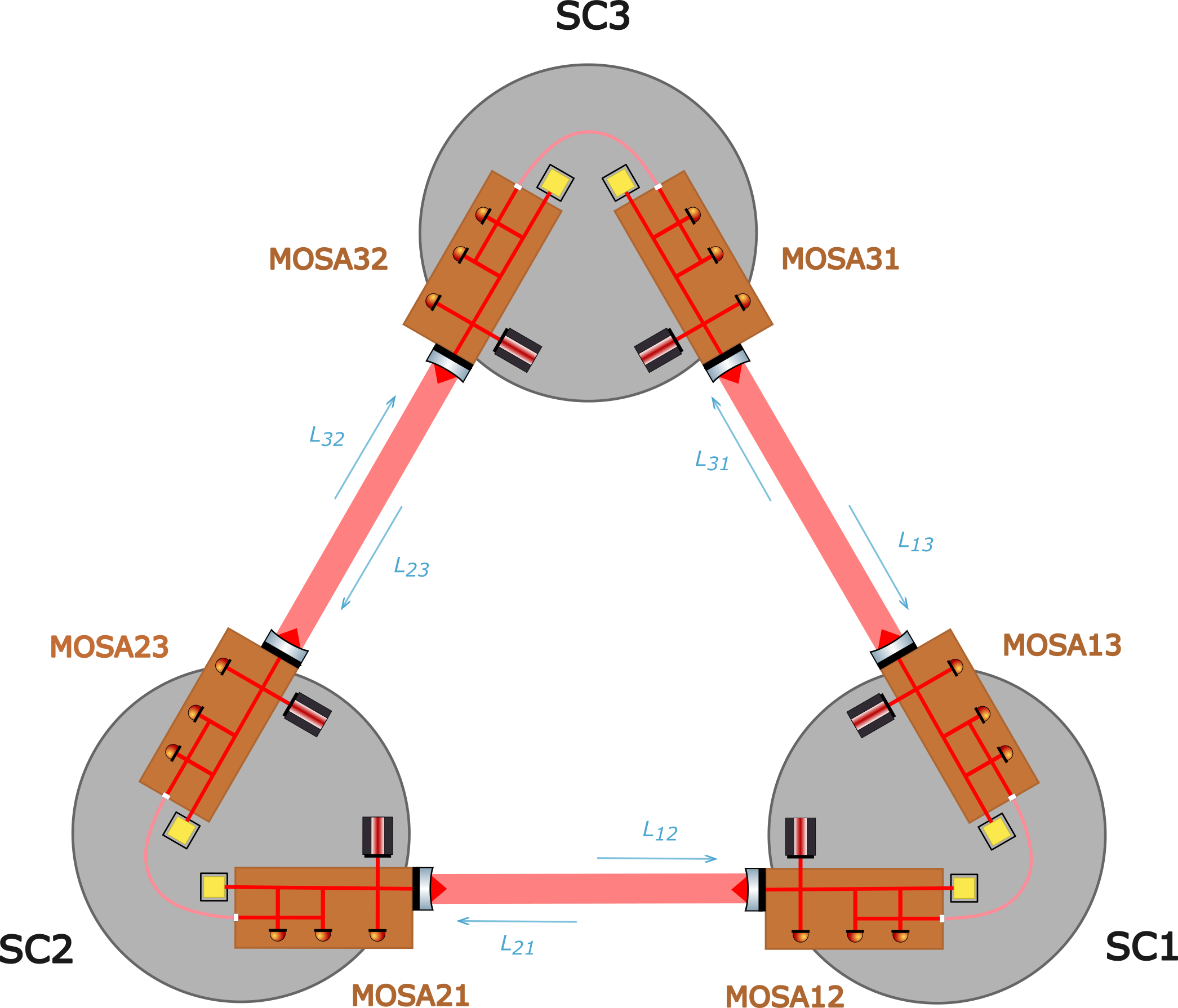}
  \caption{Schematic diagram of Taiji constellation.}
  \label{fig:tj_cnstl}
\end{figure}

As we have mentioned before, Taiji consists of three SCs around the Sun.
Each SC carries two movable optical sub-assemblies (MOSAs) which functions as transceiver and measurement unit of interferometry. 
Depicted in \Fig{fig:tj_cnstl}, we adopt the conventional indexing scheme \cite{Du:2025xdq,PhysRevD.103.123027} for facilities and observables: 
three SCs are labeled from $1$ to $3$ 
in a direction opposite to the cartwheel rotation.
MOSAs and laser paths are labeled with two indices $ij$, where $i,\,j$ denote receiving and emitting SCs, repectively.
Raw measurements are labeled according to the MOSA or SC on which they are performed.

At present, our primitive objective is to establish a basic pipeline for the detection of SGWB.
Consequently, we will only focus on TDC II training sets \cite{tdc2training}. 
The data generation processes for these training sets are simplified, 
aside from the provided parameters of injected GW signals, fewer physical effects are modeled.
In terms of noise, two main sources of noise are considered: 
test-mass acceleration (ACC) noise arises from deviations of the test-mass from ideal geodesic motion. 
Optical metrology noise (OMS) encompasses all the contributions from optical readout system.
The PSDs are given by
\begin{equation}
  \begin{aligned}
    S_\text{acc}(f) &=  A_\text{acc}^2
        \left[1 + \left(\frac{0.4\,\text{mHz}}{f}\right)^2 \right]
        \left[1 + \left(\frac{f}{8\,\text{mHz}} \right)^4 \right]\left(\frac1{2\pi fc}\right)^2, \\
    S_\text{oms}(f) &= A_\text{oms}^2
          \left[1 + \left(\frac{2\,\text{mHz}}{f}\right)^4 \right]\left(\frac{2\pi f}c\right)^2. \\
  \end{aligned}
\end{equation}
The nominal amplitudes are $A_\text{acc}=3\,\text{fm}/\text{s}^2/\sqrt{\text{Hz}}$ and $A_\text{oms}=8\,\text{pm}/\sqrt{\text{Hz}}$.
While in the process of simulation, the noises are uncorrelated, but not identically distributed in each MOSA.
The actual noise amplitudes of facilities in individual MOSAs are determined by applying $\pm20\%$ fluctuations around the nominal value.

Space-borne detectors measure the change of distance between free falling test-masses located on seperated SCs through laser interferometry.
But in practice, this is achieved indirectly by combining multiple interferometric signals
and constructing six intermediate $\eta$ observables in the preprocessing stage \cite{PhysRevD.98.042003}. At this level, noise from motion of optical bench
and three degree of freedom of laser are canceled. However, under the current limitation of laser stability,
$\eta$ observables remain dominant by laser phase noise.
This kind of noise is further suppressed by the time delay interferometry (TDI) \cite{PhysRevD.59.102003,PhysRevD.68.061303,Tinto:2004wu}.
TDI observable is formed by the linear combinations
of the time delayed $\eta$ observables:
\begin{equation}
  \Delta = \sum_{ij}\mathcal{P}_{ij}\eta_{ij},
\end{equation}
where $\mathcal{P}_{ij}$ is a polynomial of delay operator $\mathcal{D}_{ij}$. 
Details about the delay operator and the specific TDI combinations of this work are provided in \Appd{sec:TDI}. 
For the Taiji mission, the second generation of TDI is adopted. 
The training sets comprise Michelson type TDI observables $X,\,Y,\,Z$ lasting for one year.

TDC II  2.8, 2.8\_EQ, 2.9 and 2.9\_EQ are dedicated to serving the science of SGWB.
The EQ suffix denotes an analytical equal-arm orbit.
For a stationary, isotropic and unpolarized SGWB, the cross correlation spectrum is
\begin{equation}
  \langle h_A(f,\vecn)h_{A^\prime}(f^\prime,\vecn^\prime) \rangle=\frac{1}{16\pi}S_h(f)\delta_{AA^\prime}
  \delta(f-f^\prime)\delta^2(\vecn,\vecn^\prime).
\end{equation}
$S_h(f)$ can be regarded as the strain PSD of background. It is related to the fractional energy density spectrum via
\begin{equation}
  S_h(f)=\frac{3H_0^2}{2\pi^2}\frac{\Omega_\gw(f)}{f^3}, 
\end{equation}
and
\begin{equation}
  \Omega_\gw(f)=\frac{8\pi G}{3c^2H_0^2}\frac{\dif\rho_\gw}{\dif\ln f}.
\end{equation}

2.8 dataset models the astrophysical background by a power law
\begin{equation}
  \Omega_\gw(f)=A_\mathrm{ap}\left(\frac{f}{1\,\mathrm{mHz}}\right)^{\gamma_\mathrm{ap}}.
  \label{eq:omega_ap}
\end{equation}
While 2.9 models the cosmological background from PT by double broken power law model
\begin{equation}
  \Omega_\gw(f)=A_\mathrm{pt}s^9 \left(\frac{1+r_b^4}{r_b^4+s^4}\right)^{\frac{9-b}{4}}
  \left(\frac{b+4}{b+4-m+ms^2}\right)^\frac{b+4}{2},
\end{equation}
where
\begin{equation}
  m=\frac{9r_b^4+b}{r_b^4+1},\quad s=\frac{f}{f_\mathrm{pt}}.
\end{equation}
We list the parameters of spectra and selected true values in \Table{tab:truevalue}.

\section{SGWB identification}
We assume that both SGWB and noise can be depicted by stationary Gaussian process. 
Consequently, for a pair of TDI observations, the cross correlation satisfies
\begin{equation}
  \langle \Delta_I(f)\Delta_J^*(f^\prime)\rangle=\frac12\delta(f-f^\prime)\left[N_{IJ}(f)+\Gamma_{IJ}(f)S_h(f)\right].
\end{equation}
Here, $\Gamma_{IJ}(f)$ denotes the overlap reduction function (or average response) \cite{Babak:2021mhe} for the $IJ$ pair.
$N_{IJ}$ counts for the cross spectral density (CSD) of noise, 
which includes contributions from ACC and OMS noise in this work. 
Based on the simplified data preprocessing procedure, the relationship between TDI noise and raw noise is given by
\begin{equation}
  \begin{aligned}
    N_{IJ}^{\mathrm{(acc)}} &= \sum_{ij}(\mathcal{P}_{ij}^I+\mathcal{P}_{ji}^I\mathcal{D}_{ji})
    (\mathcal{P}_{ij}^J+\mathcal{P}_{ji}^J\mathcal{D}_{ji})^*
    S_{ij}^{\mathrm{(acc)}}, \\
    N_{IJ}^{\mathrm{(oms)}} &= \sum_{ij}(\mathcal{P}_{ij}^I\mathcal{P}_{ij}^{J*})S^\mathrm{(oms)}_{ij}.
  \end{aligned}
  \label{eq:noise_transfer}
\end{equation}
Although in order to achieve the accuracy of eliminating laser phase noise,
the rate of change in armlength should be taken into consideration in time delay operator.
Such high accuracy is not required for analysing TDI response to GW and other kinds of noise.
When expressing \Eq{eq:noise_transfer} in frequency domain, $\mathcal{D}_{ij}$ is just
a multiplier $\mathrm{e}^{-\mathrm{i}2\pi d_{ij}}$.

\begin{figure}[ht]
  \centering
  \includegraphics[width=.6\columnwidth]{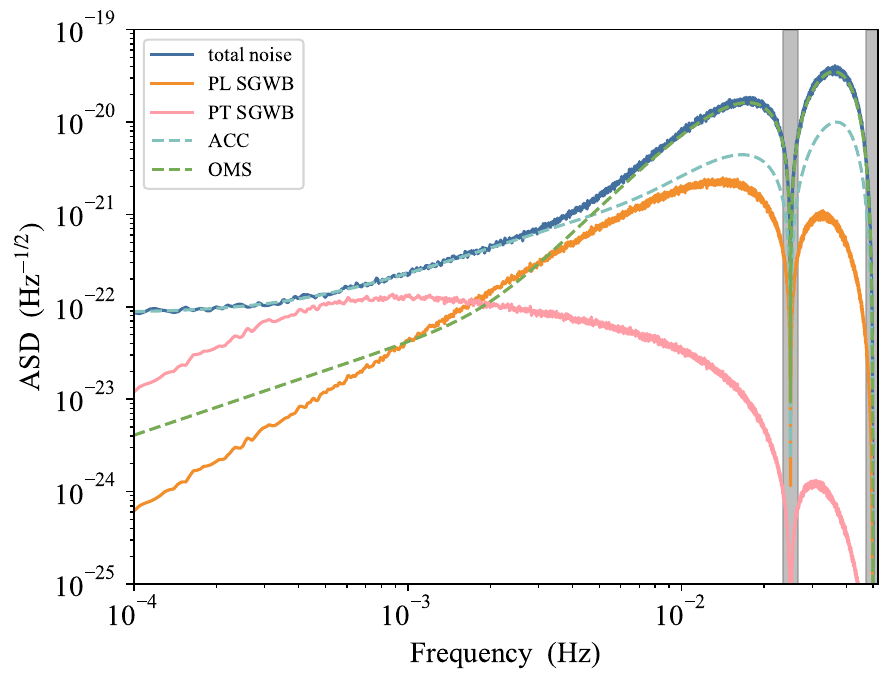}
  \caption{Amplitude spectral density (ASD) of various components in A channel. 
  All spectra except for the PT SGWB are selected from 2.8\_EQ dataset; 
  the latter are derived from 2.9\_EQ dataset. 
  Spectra estimated from observational results are plotted with solid lines, 
  while theoretically calculated spectra are shown with dashed lines.
  Grey shaded regions denote frequencies notched out in our analysis.}
  \label{fig:asd}
\end{figure}

The actual TDI combinations we adopted in the analysis are $AET$ observations, 
which derived from $XYZ$ channels with the following linear transformation:
\begin{equation}
  A=\frac{Z-X}{\sqrt{2}},\;E=\frac{X-2Y+Z}{\sqrt{6}},\;T=\frac{X+Y+Z}{\sqrt3}.
\end{equation}
In equal-arm and equal-noise constellation, $AET$ channels are orthonogonal.
That is, the covariance matrix
\begin{equation}
  C=\begin{bmatrix}\langle AA^*\rangle & \langle AE^*\rangle & \langle AT^*\rangle \\
  \langle EA^*\rangle & \langle EE^*\rangle & \langle ET^*\rangle \\
\langle TA^*\rangle & \langle TE^*\rangle & \langle TT^*\rangle \end{bmatrix},
\end{equation}
is diagonal. This character greatly simplifies the subsequent inverse process.
While neither conditions is satisfied, we retain this accustomed TDI combinations
because numerical inversion is slightly more stable.
In \Fig{fig:asd}, the PSDs of two noise components in $A$  channel are shown as dashed lines.
The ACC noise dominates at low frequencies, whereas the OMS noise is the dominant component at high frequencies.

We divide the one-year TDI observations $s(t)$ into weekly segments denoted by $s_i(t)$.
Each segment is multiplied by a Kaiser window with $\beta=14$ and then Fourier transformed into the frequency domain.
We further assume that the response pattern of detector does not vary abruptly in a segment.
Therefore, we take the configuration at intermediate moments to calculate the TDI response.
The inference of SGWB is performed with Bayesian framework. 
According to Bayes' theorem, the posterior probability satisfies
\begin{equation}
  p(k,\boldsymbol{\theta}_k|s) \propto p(s|k,\boldsymbol{\theta}_k)p(\boldsymbol{\theta}_k|k)p(k),
\end{equation}
where $p(k)$ is the prior of the model $k$, 
and $p(\boldsymbol{\theta}_k|k)$ is the prior of parameters $\boldsymbol{\theta}_k$ in this model.
In our case, the parameters to be estimated are encoded in the covariance matrix $C(f)$.
The likelihood $p(s|k,\boldsymbol{\theta}_k)$ takes the form
\begin{equation}
  \begin{gathered}
    \ln p(s|C) = -\sum_{i,f}\left[\mathrm{det}(C_i(f)) + s_i^\dagger(f)C_i^{-1}(f)s_i(f)\right],\\
    s_i(f) = \begin{bmatrix}A_{i}(f) & E_{i}(f) & T_i(f)\end{bmatrix}^T,
  \end{gathered}
  \label{eq:likelihood}
\end{equation}
where the index $i$ runs over segments and $f$ over frequency bins.
For the EQ datasets, the dependence of $C_i$ on noise and background is identical across all segments.
As a result, we can reduce the computational cost by first averaging $s_i^\dagger(f)s_i(f)$ over segments.
The resulting averaged PSDs of noise and SGWB are also shown in \Fig{fig:asd}.
Our analysis covers the frequency band from $10^{-4}$ Hz to $0.05$ Hz.
Note that a characteristic feature of the TDI algorithm is that its complete cancellation of signals in certain frequency bins,
which manifests as sharp dips of the PSDs in \Fig{fig:asd}.
We exclude frequency bins near these null frequencies.
For datasets with time-varying armlengths, pre-averaging across segments is not applicable. 
In order to speed up the computation,
we compress the data by
combining multiple adjacent frequency bins of $s_i(f)$ and averaging the estimated CSD within each combined band.
The final frequency resolution after this binning is $1\times 10^{-5}$ Hz.

\begin{table}[ht]
  \tabcolsep=.3cm
  \centering
  \begin{tabular}{cccc}
    \hline\hline
    Datasets & Parameters & Injected values & Priors \\
    \hline
    \multirow{2}*{2.8} &
    $A_\mathrm{ap}$ & $1\times10^{-12}$  & $\mathcal{U}[10^{-14},\, 5\times10^{-11}]$ \\
    & $\gamma_\mathrm{ap}$ & $2/3$ & $\mathcal{U}[-1,\, 2]$\\
    \hline
    \multirow{4}*{2.9} &
    $A_\mathrm{pt}$ & $1\times10^{-10}$ & $\mathcal{U}[10^{-14},\, 5\times10^{-9}]$\\
    & $f_\mathrm{pt}$ (mHz) & $0.2$ & $\mathcal{U}[0.1,\, 0.5]$\\
    & $r_b$ & $0.15$ & $\mathcal{U}[0.02,\, 0.5]$\\
    & $b$ & $0.65$ & $\mathcal{U}[0.2,\, 1.5]$\\
    \hline\hline
  \end{tabular}
  \caption{Parameters of SGWB spectra and priors used for Bayesian analysis in TDC II training datasets.
  EQ datasets share identical injected values with the realistic orbit datasets.}
  \label{tab:truevalue}
\end{table}
We adopt uniform priors to all free parameters in the template-based analysis.
For the amplitude of noise, we use a uniform prior spanning $\pm25\%$ fluctuations around the nominal values.
Priors of physical parameters that determine the spectrum of SGWB are listed in \Table{tab:truevalue}.
MCMC sampling is performed using the \texttt{Eryn} \cite{Karnesis2023,michael_katz_2023_7705496} package.
The default affine invariant stretch-move proposal \cite{ForemanMackey2012emceeTM} is well-suited for our application.

For the TDC II data challenge, the power spectra of both noise and injected SGWB are known for users.
In a realistic observational scenario, however, this information is unavailable. 
While the noise properties can be partly informed by ground-based experiments,
the spectral density of SGWB remains largely unknown.
Although the astrophysical background from CBCs is widely assumed to follow a power law with spectral index $\gamma_\mathrm{ap}=2/3$, 
the cosmological component is difficult to predict across the entire Taiji frequency band.
Moreover, the superposition of all potential contributions result in a composite spectrum of greater complexity and uncertainty.
Therefore, we prefer to retain flexibility in modeling the SGWB without relying on a precise theoretical template.
A generic approach to implement this idea is interpolation. We fit the spectrum curve
in the $\log\Omega_\gw$--$\log f$ plane using a piecewise quadratic spline function.
By adjusting the knot positions, such an interpolation can adapt to any continuous and smooth spectral shape.
A key question, however, is how to choose the number of knots.
Trans-dimensional methods are particularly suited for this purpose, 
as they simultaneously infer the optimal model and the corresponding parameters.
We therefore employ the reversible jump MCMC (RJMCMC) \cite{biomet82.4.711} to estimate the spectrum.

In a trans-dimensional problem, the parameter space $\boldsymbol{\theta}_k$ has different dimensions for different model $k$.
To transit to a new model from $k$, a candidate model $k^\prime$ is proposed with probability $j(k^\prime|k)$,
and random variables $\boldsymbol{u}$ are drawn from a designed proposal
$q(\boldsymbol{u}|\boldsymbol{\theta}_{k}, k)$. The parameters of new state are obtained via 
$(\boldsymbol{\theta}^\prime_{k^\prime},\boldsymbol{u}^\prime)=g(\boldsymbol{\theta}_k,\boldsymbol{u})$,
where $g$ is a specified diﬀeomorphism. Like Metropolis-Hasting algorithm, this new state is accepted with probability
\begin{equation}
  \min\left(1, \frac{p(k^\prime,\boldsymbol{\theta}^\prime_{k^\prime}|s)j(k|k^\prime) q(\boldsymbol{u}^\prime|\boldsymbol{\theta}^\prime_{k^\prime}, k^\prime)}
  {p(k,\boldsymbol{\theta}_k|s)j(k^\prime|k) q(\boldsymbol{u}|\boldsymbol{\theta}_{k}, k)}
  \left|\frac{\partial(\boldsymbol{\theta}^\prime_{k^\prime},\boldsymbol{u}^\prime)}{\partial(\boldsymbol{\theta}_k,\boldsymbol{u})}\right|\right)
  \label{eq:rj_alpha}
\end{equation}
In our implementation, the model index $k$ corresponds to the number of knots, and their positions form the parameter space $\boldsymbol{\theta}_k$.
Trans-dimensional sampling is typically regarded as challenging and often requires careful tuning of 
the proposal mechanisms to achieve satisfactory performance.
Fortunately, our model belongs to the nested category. When jumping to higher-dimensional space, we can add a new knot while keeping the existed knots unchanged.
Consequently, $g$ reduces to the identity transformation and the Jacobian in \Eq{eq:rj_alpha} become unity.

We fix the two edge knots to the boundaries of our analysis frequency band.
The prior for the positions of the interior knots is uniform distribution within a rectangle region in $\log\Omega_\gw-\log f$ plane,
with $\log\Omega_\gw$ constrained to $[-14,\,-9]$.
The maximum numbers of interior knots is set to $16$.
The default RJMCMC sampler also proposes new knots with this prior.
Although the results in the next section show that the setup is workable now, we note that this may lead to inefficient sampling for a more complex problem.
Additionally, the stretch proposal can not be directly applied to trans-dimensional steps. 
We therefore enable the gibbs sampling, 
preserving the stretch move to noise parameters and use a Gaussian proposal to in-model update of the knots positions.

\section{Results}
\begin{figure}[ht]
  \centering
  \includegraphics[width=.6\columnwidth]{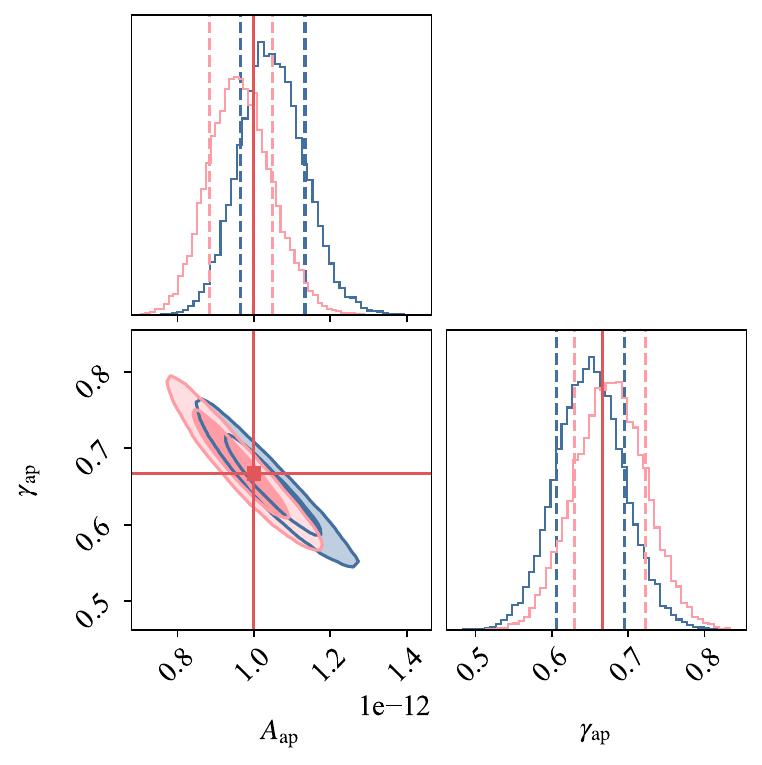}
  \caption{Posterior distributions for the parameters of the astrophysical SGWB from in 2.8\_EQ (blue) and 2.8 (pink) datasets. Red lines indicate the injected values and vertical dashed lines mark the $1\sigma$ CIs.
  The shaded regions represent the $68\%$ and $95\%$ confidence levels.}
  \label{fig:pe_sgwb_28t}
\end{figure}

\begin{figure}[ht]
  \centering
  \includegraphics[width=.6\columnwidth]{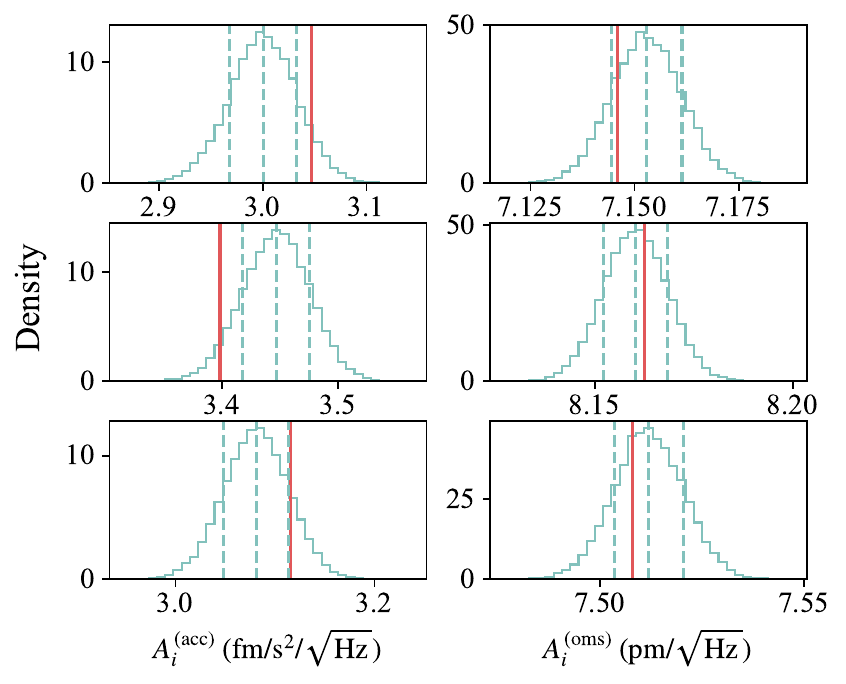}
  \caption{Posterior distributions for ACC and OMS noise amplitudes in 2.8\_EQ datasets,
  arranged in ascending order of SC index $i$ from top to bottom.
  The red lines denote injected values while the vertical dashed lines show the median and $1\sigma$ CI.}
  \label{fig:pe_noise_28EQt}
\end{figure}

As an initial validation, we first demonstrate the recovery of astrophysical SGWB in 2.8\_EQ dataset using the template-based method.
In \Fig{fig:pe_sgwb_28t}, the posterior distribution is shown in blue shaded region. Both the amplitude and the power index of the injected astrophysical SGWB are successfully recovered within the $1\sigma$ credible interval (CI). 
This result provides evidence for the applicability of our pipeline.
As note in \cite{PhysRevD.82.022002}, the amplitudes of individual noise components exhibit a strong degeneracy in the equal-arm orbital configuration.
We define an averaged amplitude for each SC over its associated MOSAs as $A_i=(A_{ij}+A_{ik})/2$.
We find that it can mitigate the redundant degrees of freedom. 
Hence, we present the posterior distribution of noise by this averaged amplitudes in \Fig{fig:pe_noise_28EQt}.
The error in estimating the amplitude of low-frequency ACC noise is slightly larger, but still within the $2\sigma$ CI.

\begin{figure}[ht]
  \centering
  \includegraphics[width=.6\columnwidth]{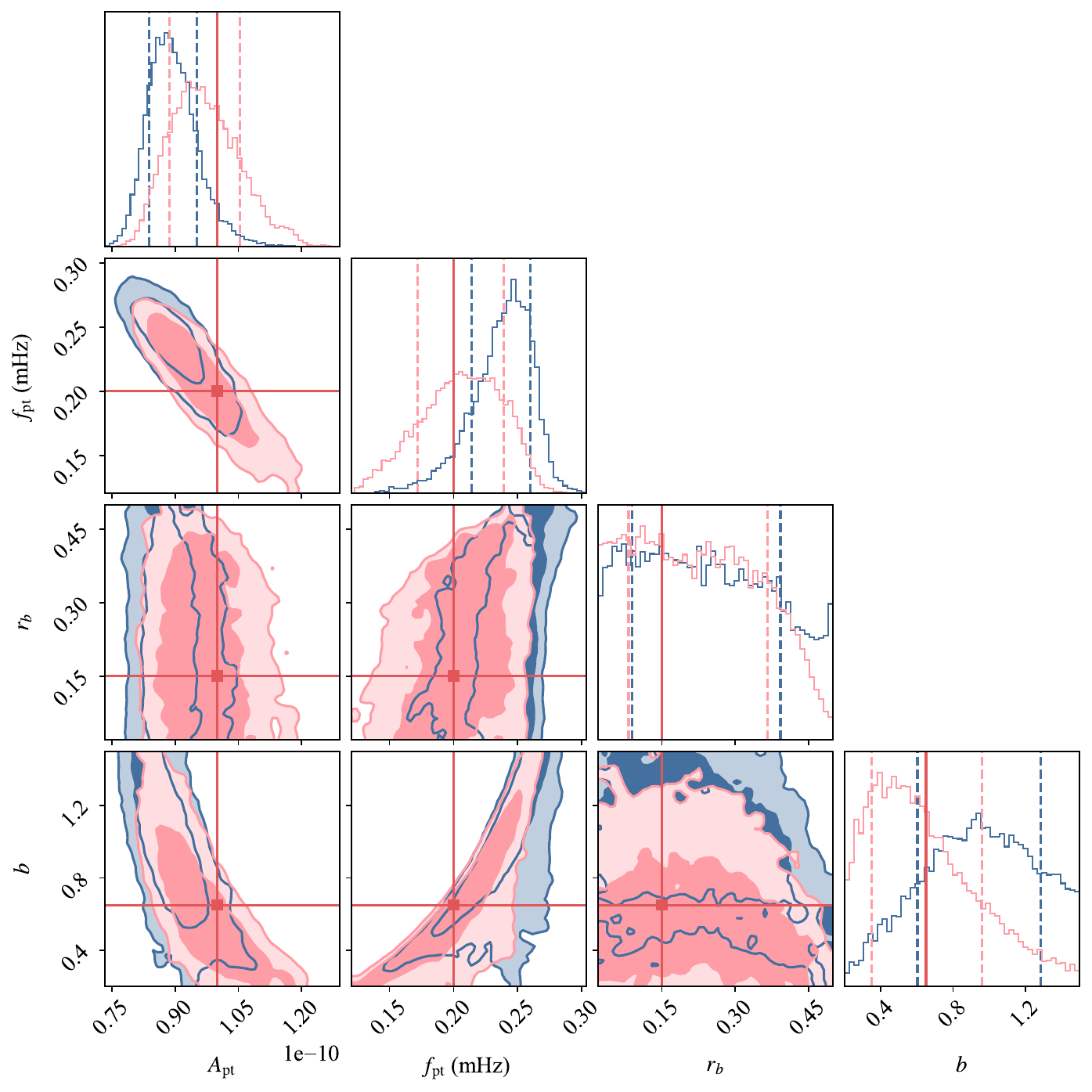}
  \caption{Posterior distributions for the parameters of the PT SGWB from in 2.9\_EQ (blue) and 2.9 (pink) datasets. Red lines indicate the injected values and vertical dashed lines mark the $1\sigma$ CIs.
  The shaded regions represent the $68\%$ and $95\%$ confidence levels.}
  \label{fig:pe_sgwb_29t}
\end{figure}

Afterwards, we will focus on resolving the SGWB in the flexing-arm scenario, putting results of noise parameters to \Appd{sec:noise_params}.
In the equal-arm case, the T channel constitutes a null channel insensitive to the GW signals, 
making it useful for monitoring instrumental noise.
However, its responses to noise and GW are significantly affected by variations of armlength \cite{Hartwig:2023pft}. 
We have done an attempt, that is, recovering astrophysical SGWB using an equal-arm model on flexing-arm dataset.
The results show a strongly biased estimate: 
$95\%$ CI for $A_\mathrm{ap}$ is $[1.66,\,1.72]\times10^{-12}$
and the posterior for $\gamma_\mathrm{ap}$ concentrated at $-1$, which is the lower bound of our chosen prior.
This confirms that the equal-arm model is unsuitable for realistic orbit configuration.
In constrast, \Fig{fig:pe_sgwb_28t} also shows the posterior distributions for astrophysical SGWB parameters after flexing-arm effect considered.
The injected values are successfully recovered and 
the variance of armlength does not bring significant changes to the confidence contours. 
A similar analysis to cosmological PT SGWB is presented in \Fig{fig:pe_sgwb_29t}.
We have obtained meaningful constraints on the background intensity and its peak frequency,
while the constraint on the parameter $r_b$ remains relatively weak.
Our analysis pipeline takes about $\mathcal{O}(10)$ hours using 64 cores of an AMD EPYC chip.

\begin{figure}[ht]
  \centering
  \includegraphics[width=.6\columnwidth]{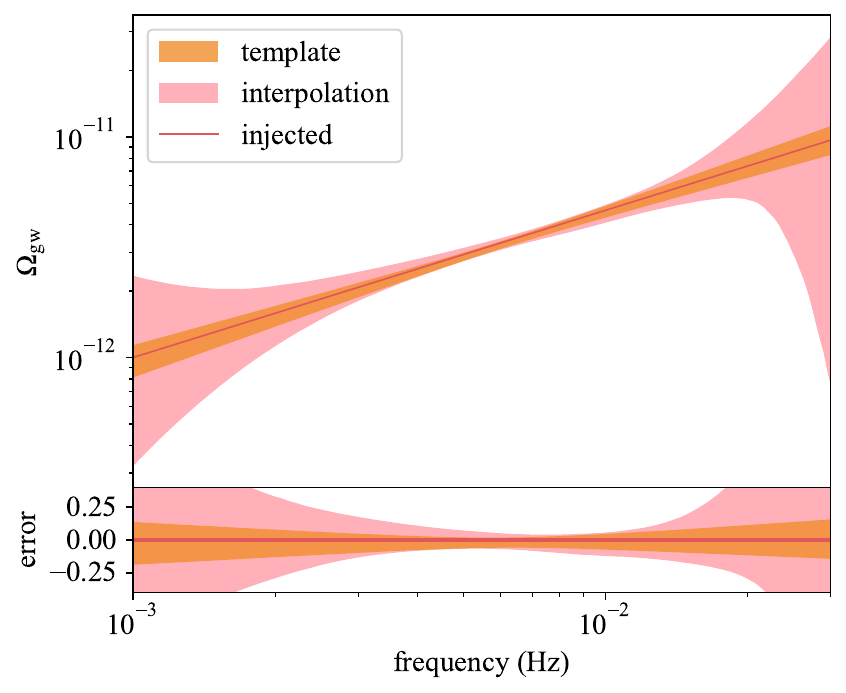}
  \caption{Top: recovered spectrum of astrophysical SGWB in 2.8 dataset.
  The red lines denote injected spectrum while the shaded regions represent $95\%$ CIs.
  Bottom: the corresponding relative errors of $95\%$ CIs.}
  \label{fig:pe_omega_28}
\end{figure}

\begin{figure}[ht]
  \centering
  \includegraphics[width=.6\columnwidth]{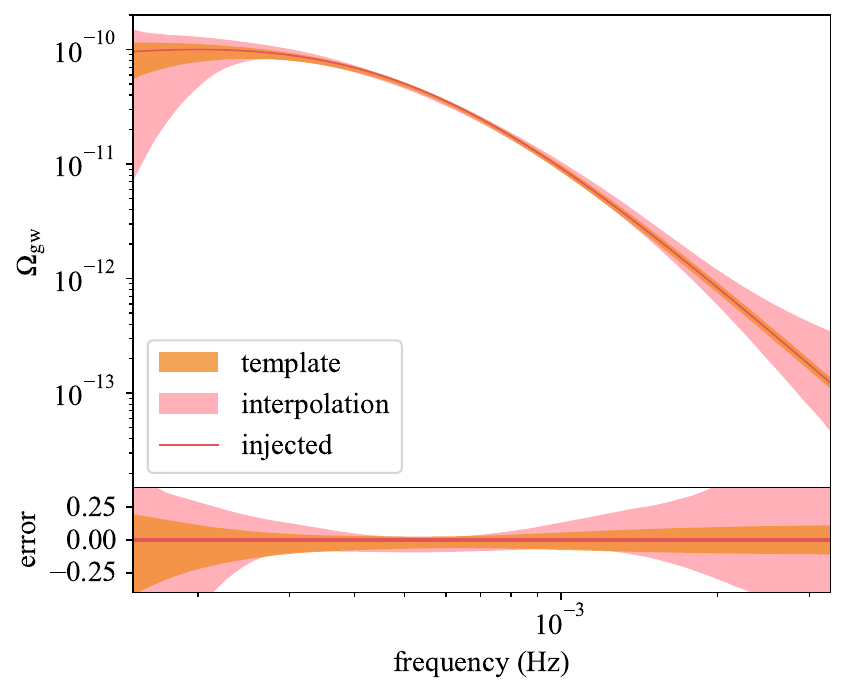}
  \caption{Top: recovered spectrum of PT SGWB in 2.9 dataset.
  The red lines denote injected spectrum while the shaded regions represent $95\%$ CIs.
  Bottom: the corresponding relative errors of $95\%$ CIs.}
  \label{fig:pe_omega_29}
\end{figure}

Next, we present the recovery of the energy-density spectrum based on the RJMCMC interpolation method. Compared to the template-based evaluation, we find that the proposal and mixing between MCMC chains become more challenging. 
Consequently, we double the number of burn‑in samples and increase the thinning interval of the MCMC chain. 
Gibbs sampling also introduces additional computational cost. It takes about 50 hours with the same number of cores.
Each sample from this sampler consists of a set of knots, which corresponds to one possible spectrum.
To translate these samples into constraints on the SGWB, we compute the corresponding $\Omega_\gw(f)$ for all samples 
and then derive the CI at selected frequencies. The results are shown in \Fig{fig:pe_omega_28} and \Fig{fig:pe_omega_29}.
For comparison, the recovered spectrum from the template-based method is also shown.
Noticing that the $1$ mHz pivot frequency in \Eq{eq:omega_ap} is below the sensitive frequency band of the detector.
This naturally leads to the correlation of the posterior between $A_\mathrm{ap}$ and $\gamma_\mathrm{ap}$ in \Fig{fig:pe_sgwb_28t}.
Both methods successfully reconstruct the injected spectra. 
Within the detector's sensitivity band, the two results show a high degree of consistency,
with nearly identical uncertainty range. 
For known template and interpolation methods, in the most sensitive frequency band, a $95\%$ CI corresponds to approximately $8\%$ and $10\%$ errors, respectively.
While the parameterized template can extrapolate the spectrum beyond the sensitive band,
such extrapolation is not informed by the data themselves.
In contrast, the interpolation spectrum reflects this lack of information by showing rapidly growing uncertainties in those regions.

\section{Conclusions \& Future work}
We have presented a preliminary framework for detecting SGWB for Taiji mission.
The code usd in this work is publicly available at \url{https://github.com/DrizzleatDusk/TaijiSGWB}.
The present work focuses on the official Taiji training datasets,
which incorporate realistic features such as unequal noise levels and time-varying armlengths.
We show that applying an equal-arm model to the flexing-arm datasets introduces significant bias.
To address this, we segment time-domain strain data and compute the detector response individually for
each segment, subsequently evaluating the likelihood via inversion of the full, non-diagonal covariance matrix.

Our analysis covers the frequency band from $10^{-4}\sim 0.05$ Hz. By using template-based method, 
we confirm that Taiji's projected sensitivity to detect both an astrophysical SGWB with
$\Omega\sim1\times10^{-12}$ at 1 mHz and a PT SGWB with 
$\Omega\sim1\times10^{-10}$ at 0.2 mHz.
In practice, however, the precise spectral shape of the SGWB may be unknown.
To overcome the limitation of template-based method, we employ a flexible, model-independent approach
in which the SGWB spectrum is reconstructed with a quadratic interpolation.
Since the optimal number and locations of spline knots are not known in advance, 
we sample them simultaneously using a trans-dimensional MCMC algorithm.
This method is shown to successfully reconstruct the injected background spectrum without bias.

A key limitation of the current work is the complete omission of GB foreground.
Therefore, we cannot yet say that the analysis is fully satisfied the objectives of the training datasets. 
The immediate next step is to incorporate the GB foreground.
As mentioned earlier, the CGB is anisotropic and exhibits cyclostationary modulation in time domain. 
While phenomenological treatments are possible \cite{Boileau:2021sni}, 
a more principled approach would be to model the response of Taiji to an anisotropic background,
since the modulation pattern is intrinsically linked to the spatial distribution of GBs in the Milky Way \cite{Criswell:2024hfn}.
Work in this direction is currently underway.

\section*{Acknowledgments}
The authors thank Minghui Du for the guidance on the \texttt{Triangle} toolkit
and support of data simulation.
Yang would also like to thank Jingyi Wu for the discussions during the course.
YJ is supported by the China Postdoctoral Science Foundation under Grant Number 2025M783376. QGH is supported by the grants from NSFC (Grant No.~12547110, 12475065, 12447101) and the China Manned Space Program with grant no. CMS-CSST-2025-A01.

\appendix
\section{TDI}
\label{sec:TDI}
TDI is a process that cancel the significant laser phase noise and forming a virtual laser interference signal.
Delay operator $\mathcal{D}_{ij}$ is defined as 
\begin{equation}
  \mathcal{D}_{ij}f(t)= [1-\dot{d}_{ij}(t)] f[t-d_{ij}(t)],
\end{equation}
where $f(t)$ is an arbitrary function defined in time domain and 
$d_{ij}$ is light travel time along $L_{ij}$.
If the rate of change in armlength $\dot{d}_{ij}$ is negligible, 
the delay operator, as its name suggests, acts as a time delay transformation.

The second generation of TDI is necessary for Taiji mission,
Michelson $X$ TDI is constructed by the following polynomial operators:
\begin{equation}
  \begin{aligned}
    \mathcal{P}_{12} &= 1-\mathcal{D}_{131}-\mathcal{D}_{13121}+\mathcal{D}_{1213131},\\
    \mathcal{P}_{23} &= 0,\\
    \mathcal{P}_{31} &= -\mathcal{D}_{13}+\mathcal{D}_{1213}+\mathcal{D}_{121313}-\mathcal{D}_{13121213},\\
    \mathcal{P}_{21} &= \mathcal{D}_{12}-\mathcal{D}_{1312}-\mathcal{D}_{131212}+\mathcal{D}_{12131312},\\
    \mathcal{P}_{32} &= 0,\\
    \mathcal{P}_{13} &= -1 + \mathcal{D}_{121}+\mathcal{D}_{12131}-\mathcal{D}_{1312121}.
  \end{aligned}
  \label{eq:tdiX}
\end{equation}
Where $\mathrm{D}_{ijk\cdots}$ is an abbreviation for the continuous action of the delay operator
$\mathcal{D}_{ij}\mathcal{D}_{jk}\cdots$.
$Y$ and $Z$ channels can be constructed by cyclic the index in \Eq{eq:tdiX}.

\section{Noise amplitudes}
\label{sec:noise_params}
\begin{figure}[ht]
  \centering
  \includegraphics[width=.6\columnwidth]{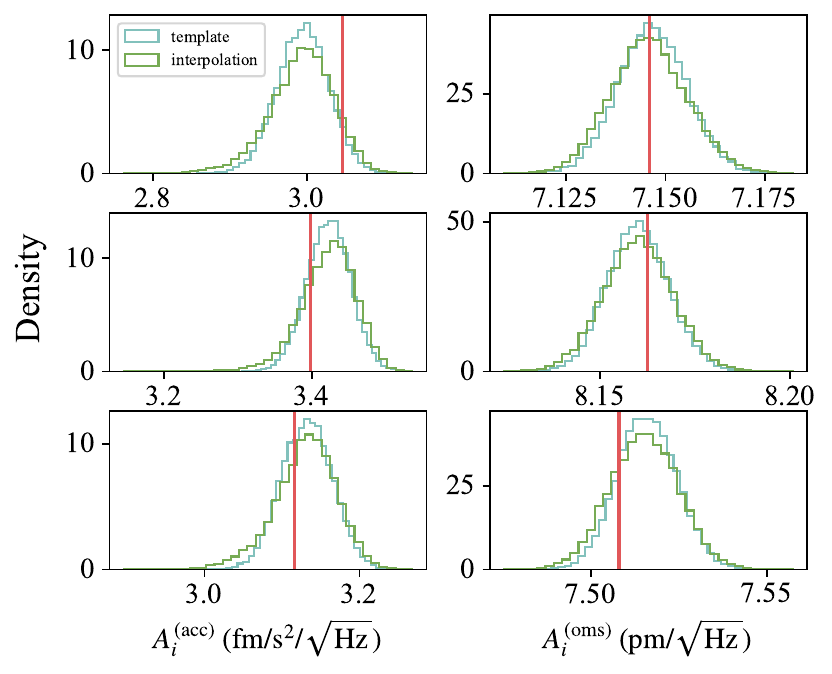}
  \caption{Posterior distributions for ACC and OMS noise amplitudes in 2.8 datasets,
  arranged in ascending order of SC index $i$ from top to bottom.
  The red lines denote injected values.}
  \label{fig:pe_noise_28}
\end{figure}

\begin{figure}[ht]
  \centering
  \includegraphics[width=.6\columnwidth]{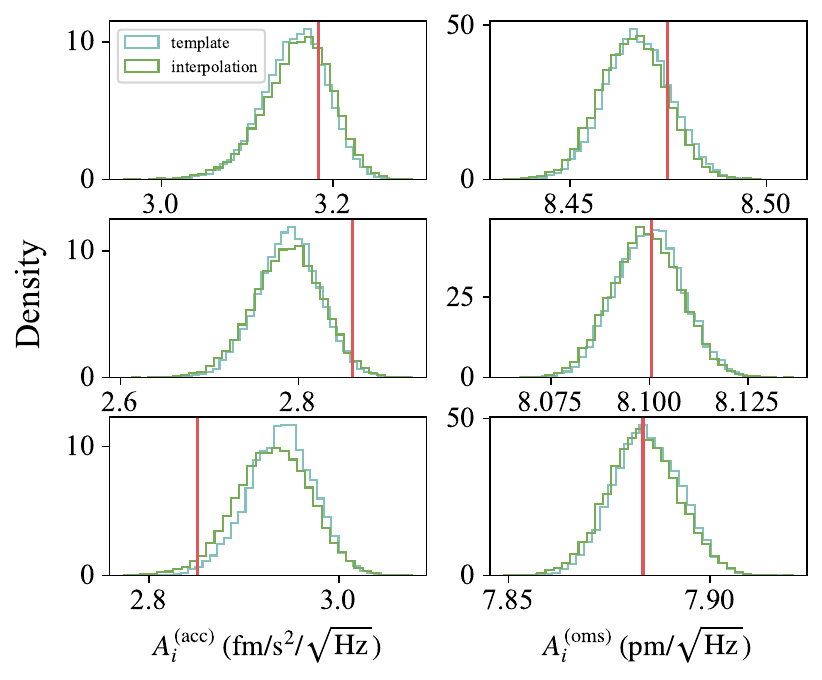}
  \caption{Posterior distributions for ACC and OMS noise amplitudes in 2.9 datasets,
  arranged in ascending order of SC index $i$ from top to bottom.
  The red lines denote injected values.}
  \label{fig:pe_noise_29}
\end{figure}

We supplement the posteriors of noise amplitude here.
\Fig{fig:pe_noise_28} and \Fig{fig:pe_noise_29} denote the distribution of noise amplitudes for 2.8 and 2.9 datasets, respectively.

\bibliography{refs.bib}

\end{document}